\documentclass[12pt,preprint]{aastex}

\shorttitle{J1420--0545: The largest giant radio galaxy}
\shortauthors{Machalski et al.}

\begin{document}

\title{J1420--0545: The radio galaxy larger than 3C236}

\author{J. Machalski$^1$\email{machalsk@oa.uj.edu.pl} D.
Kozie{\l}-Wierzbowska$^1$\email{eddie@oa.uj.edu.pl} 
M. Jamrozy$^1$\email{jamrozy@oa.uj.edu.pl} D.J. Saikia$^2$\email{djs@ncra.tifr.res.in}}
\affil{$^1$ Astronomical Observatory, Jagellonian University,
ul. Orla 171, PL-30244 Krakow, Poland}
\affil{$^2$ National Centre for Radio Astrophysics, Tata Institute of Fundamental Research (TIFR), Pune University Campus, Post Bag 3,
Pune 411 007, India}

\begin{abstract}
We report the discovery of the largest giant radio galaxy, J1420-0545: 
a FR type II radio source with an angular size of 17.4' identified
with an optical galaxy at $z$=0.3067. Thus, the projected linear size of the radio
structure is 4.69 Mpc (if we assume that $H_{0}$=71 km\,s$^{-1}$Mpc$^{-1}$, $\Omega_{\rm m}$=0.27,
and $\Omega_{\Lambda}$=0.73). This makes it larger than 3C236, which is the largest double
radio source known to date. New radio observations with the 100 m Effelsberg telescope 
and the Giant Metrewave Radio Telescope, as well as optical identification with
a host galaxy and its optical spectroscopy with the William Herschel Telescope 
are reported. The spectrum of
J1420$-$0545 is typical of elliptical galaxies in which continuum emission with the
characteristic 4000\,\AA\hspace{1mm}discontinuity and the $H$ and $K$ absorption lines are
dominated by evolved stars. The dynamical age of the source, its jets' power, the energy density, and the
equipartition magnetic field are calculated and compared with the corresponding
parameters of other giant and normal-sized radio galaxies from a
comparison sample. The source is characterized by the exceptionally low density
of the surrounding IGM and an unexpectedly high expansion speed of the source
along the jet axis. All of these may suggest a large inhomogeneity of the IGM.

\end{abstract}


\section{Introduction}

The existence of bright hot spots at the edges of most of FR type II radio sources indicates
that the jets ejected from a "central engine" in the AGN encounter resistance to their
propagation. These supersonically advancing hotspots are likely to be confined 
by the ram pressure of
the external environment, which on the largest scales is the intergalactic medium (IGM).
The shocked jet material and shocked IGM form a cocoon surrounding the pair
of jets. In a number of FR type II radio sources, such a cocoon is observed as
a low-brightness bridge between the radio lobes \citep{leahy}.
While the lengthening of the cocoon is governed by the balance between the
jet's thrust and the ram pressure of the IGM, the width of the cocoon is
determined by its mean internal pressure, which is on the order of the kinetic energy
delivered by the jets divided by the cocoon's volume. As both of these factors
are functions of time, it is possible that the lateral expansion of the
cocoon could be faster than the rate at which the jets delivered energy. Thus, the
cocoon's pressure would decrease until it reached an equilibrium with the IGM
pressure. However, in a number of papers (e.g. \citealt{begel}; \citealt{daly};
\citealt{wdw}) there are arguments that the cocoons in many
observed radio sources are still overpressured with respect to the IGM. Only
the very tenuous material in the bridges of the oldest and largest sources
is expected to have attained such an equilibrium state, in which the radiating
particles and the magnetic field are balanced, and the pressure of 
the relativistic plasma equals the pressure of the gaseous environment (cf. \citealt{nath}).
Consequently, a determination of the physical conditions in these diffuse
bridges of large-size double radio sources (as they are far away from the ram pressure-confined heads
of the source) gives us an independent tool with which to probe the pressure of the IGM.

The above approach was applied by us \citep{mkj} to a sample of giant radio galaxy (GRG) 
candidates located in the southern sky hemisphere with the aim of studying properties 
of the cosmological evolution of the IGM.
The radio galaxy discussed in this paper is a member of that sample. The source J1420$-$0545 
was discerned in the VLA 1.4 GHz surveys FIRST \citep{becker} and NVSS \citep{condon} 
as a 17.4' large FR type II radio 
structure consisting of two extended lobes with a total flux density of only 87 mJy
and a central compact core.
The core is unresolved with the 5\arcsec$\times$5\arcsec\hspace{1mm} beam in the FIRST survey
and has a flux density of 2.7 mJy. The structure is highly collinear and symmetric; the ratio
of separation between the core
and the brightest regions in the lobes is 1.08, and the misalignment angle is 1.3$\degr$. 
The FIRST map reveals also compact hot spots in both lobes, which implies strong
shocks and {\sl in situ} reacceleration of relativistic particles, finally inflating the
lobes. In order to check whether the observed lobes are or are not possibly connected by a
 bridge undetected in the NVSS because of the lack of short baselines, we made
supplementary single-dish 21 cm observations using the Effelsberg 100 m radiotelescope,
\footnote{The Effelsberg 100m telescope is operated by the Max-Planck-Institut
f\"{u}r Radioastronomie (Bonn, Germany)} as well as 50 cm observations with the Giant Metrewave
Radio Telescope (GMRT).\footnote{The GMRT is a national facility operated by the National
Centre for Radio Astrophysics of the Tata Institute of Fundamental Research (Pune, India)}
These observations, the final 1.4 GHz map of the investigated radio galaxy obtained
from a combination of the NVSS and Effelsberg data, and the 619 MHz GMRT map are presented
in \S~2. The radio
core position coincides perfectly with that of a galaxy with $R\sim$19.65 mag and $B\sim$21.82 mag at
R.A.(J2000.0) = 14$^{\rm h}$20$^{\rm m}$23.80$^{\rm s}$ and decl.(J2000.0) = $-$05\degr45\arcmin28.8\arcsec,
according to the magnitude calibration and astrometric position in the Digitized Sky Survey
(DSS) data base. This DSS red magnitude suggests a rather distant galaxy whose 
photometric redshift estimate would be between 0.46 and 0.42, according to the Hubble diagrams
for GRGs published by \citet{schoen} and \citet{lara}, respectively.
However, the standard deviations in the apparent-magnitude distributions in those diagrams for
a given redshift are about 1 mag. Therefore, supposing that the absolute magnitude
$M_{\rm R}$ of galaxies identified with very faint GRGs can be 1 $\sigma$ lower than the mean 
value, we have estimated a value of $z\approx 0.32$ for J1420$-$0545 (cf. \citealt{mkj}). The independent
photometry and optical spectroscopy of the host galaxy that confirm the above estimate are given
in \S~3, while some physical parameters and the dynamical age of the radio structure are
estimated and discussed in \S~4.  

\section{New radio observations}

\subsection{Effelsberg observations}

\begin{figure}
\epsscale{0.6}
\plotone{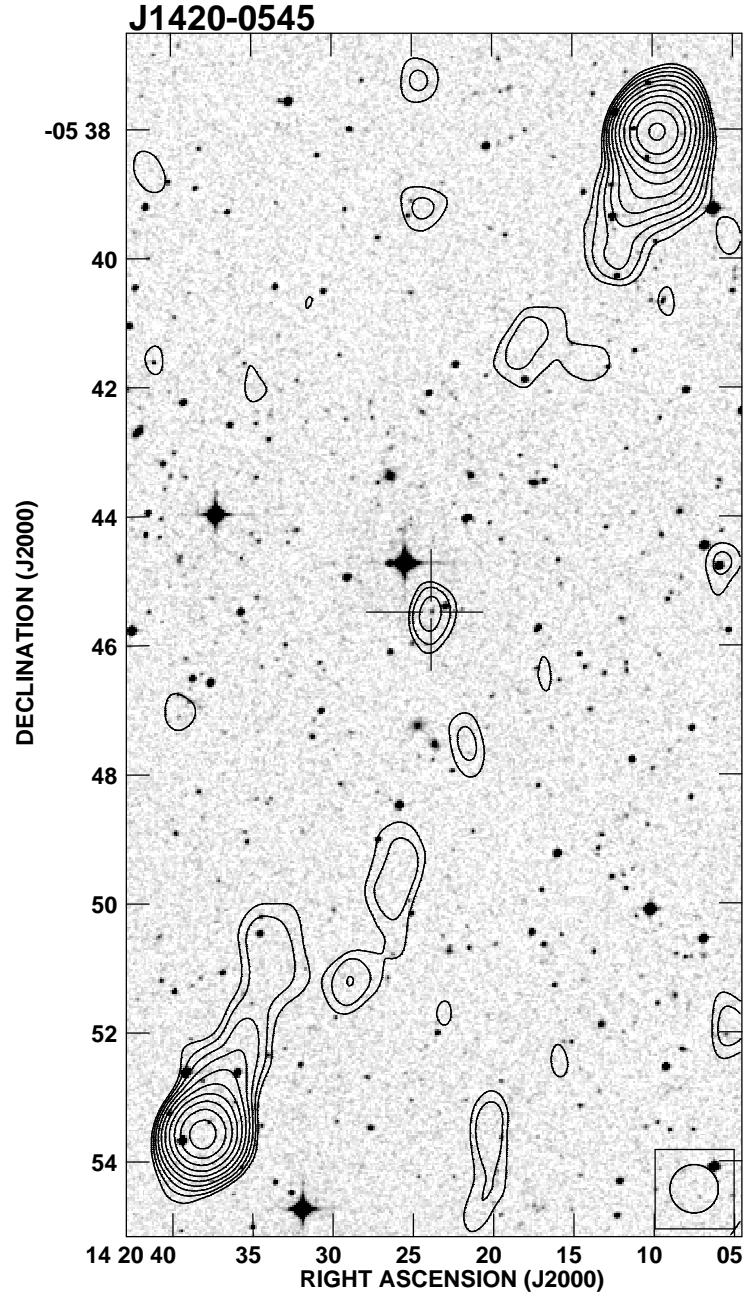}
\caption{VLA and Effelsberg contour map of J1420$-$0545 at 1.4 GHz, superposed on the optical
DSS field. Contours are logarithmic with a ratio of $2^{1/2}$ between levels; the first
contour is at 1 mJy beam$^{-1}$}.\label{fig1}
\end{figure}

The observations were carried out with the 100 m telescope on 2006 April 26--27, using the
21 cm single-horn receiver installed at the primary focus of
the antenna. The 60\arcmin$\times$60\arcmin\hspace{1mm}sky area centred at
R.A.(J2000.0)=14$^{\rm h}$20$^{\rm m}$24$^{\rm s}$ and decl.(J2000.0)=$-$05\degr45\arcmin29\arcsec
\hspace{1mm}was scanned alternately in right ascension and declination. The drive rate was
180\arcmin\hspace{1mm}per minute and the scan interval was 3\arcmin. For the flux density
calibration, the standard source 3C286 was observed. The final Effelsberg map is a combination
of the 60\arcmin$\times$60\arcmin\hspace{1mm}images, which were large enough to cover the target
source and some emission-free areas used for the zero-level determination. A combination of
the Effelsberg and NVSS images was performed using the {\sc imerg} task included in the {\tt NRAO AIPS}
package. The resulting 1.4 GHz map of the source J1420$-$0545, superposed on the optical
DSS image, is shown in Figure~\ref{fig1}. We still do not detect a bridge at this frequency.

\subsection{GMRT observations}

\begin{figure}
\epsscale{0.6}
\plotone{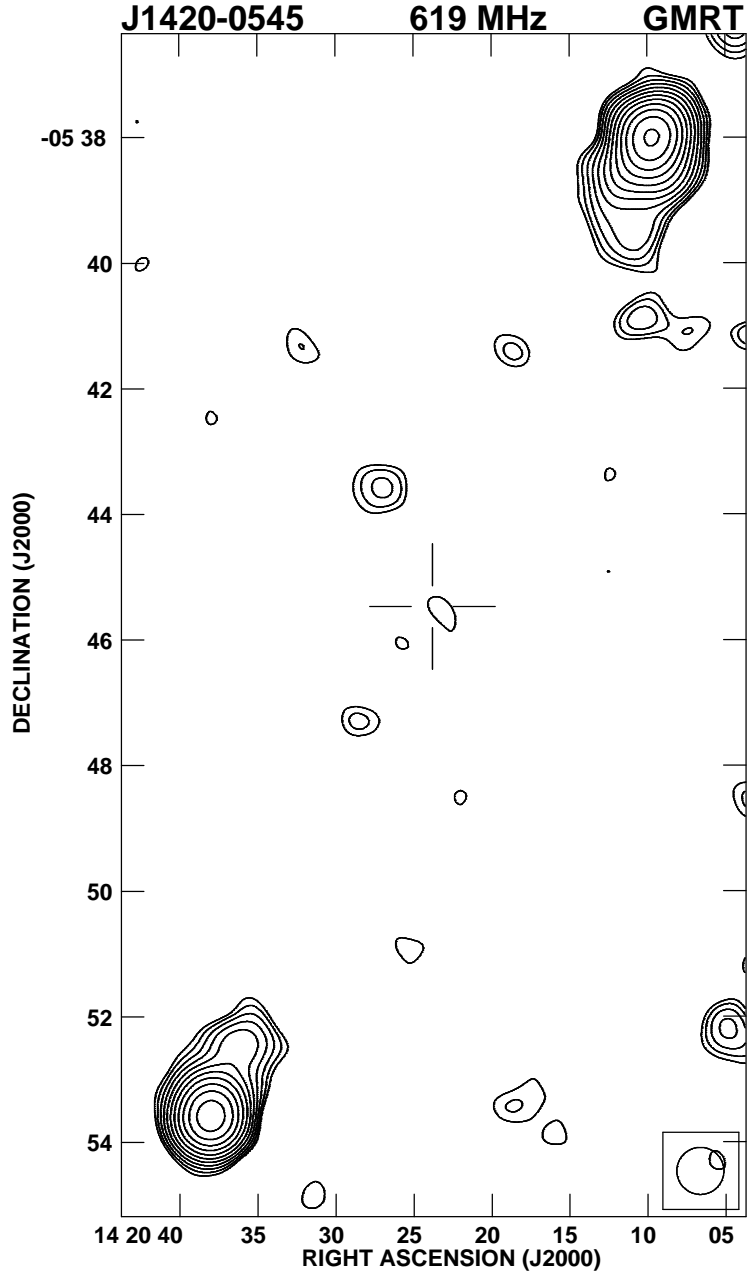}
\caption{GMRT contour map of J1420$-$0545 at 619 MHz. Contours are logarithmic with a ratio
of $2^{1/2}$ between levels; the first contour is at 1.3 mJy beam$^{-1}$}.\label{fig2}
\end{figure}

The observations of J1420$-$0545 centered at the radio core position were carried
out on 2007 November 11, in the 50 cm band. The observations were made in the standard manner,
with observations of the target source interspersed with observations of the phase calibrator  J1351$-$148.
The total observing time spent on the source was approximately 6 hr. Again the primary
flux density and bandpass calibrator was 3C286. The data were edited, calibrated, and imaged
using the standard {\tt AIPS} routines. Phase self-calibration was done for several rounds
to improve the quality of the source's image. Finally, the image was corrected for the primary
beam pattern of the antennas. The data were imaged with different angular resolutions. The
tapered 619 MHz map with a resolution of 45\arcsec$\times$45\arcsec \hspace{1mm}(i.e., with
resolution identical to that of the NVSS) and an rms noise level of about 0.5 mJy\,beam$^{-1}$
is shown in Figure~\ref{fig2}. This map also shows no bridge with a surface brightness exceeding
the above rms noise level; however, it indicates that the radio core has a very flat spectrum,
as its 619 MHz flux density is likely below 1.3 mJy.

\section{Optical observations}

 To check the DSS apparent magnitudes and to derive an absolute magnitude of the host galaxy,
an independent photometry was made at Mount Suhora Observatory with the 60 cm telescope. After
the initial standard reductions, aperture photometry was performed on the galaxy and calibration
stars using the {\tt IRAF} package {\sc apphot}. The resulting magnitudes in the $UBVRI$ photometric
system, corrected for the atmospheric extinction only (as the Galactic latitude of this galaxy
is 50\degr, the foreground Galactic extinction is close to zero), are $I$=18.44$\pm$0.08 mag,
$R$=19.65$\pm$0.08 mag, and $V$=21.0$\pm$0.2 mag. Thus, our photometry is fully compatible with
the DSS magnitudes. The combined data give an optical spectral index of $-$3.45 and an absolute
red magnitude $M_{\rm R}$ of $-$22.1$\pm$0.1, if we use the cosmological parameters given in the Abstract.

The optical spectra of the host galaxy were obtained at La Palma on 2007 August 16.
The observations conducted in the Service Programme mode were made with the ISIS double
beam spectrograph (the red and blue arms) on the 4.2 m William Herschel Telescope (WHT).\footnote{The WHT is operated by the Isaac
Newton Group in the
Spanish Observatorio del Roque de Los Muchachos of the Instituto de Astrofisica de Canarias}
The galaxy was observed in good weather and seeing conditions through a 2\arcsec \hspace{0.5mm}
wide slit. Four spectra (two per arm) were taken with the grating R316R, set up for a central wavelength
of 6447\AA\hspace{1mm}and a dispersion of 0.92\AA\hspace{1mm}pixel$^{-1}$ at the red arm, and
the grating R300B, set up for a central wavelength of 4503\AA\hspace{1mm}and a dispersion of 
0.86\AA\hspace{1mm}pixel$^{-1}$ at the blue arm. The exposure time for each spectrum was
1800 s. The spectra combined from both arms cover the wavelength range 3400\AA--8500\AA.

\begin{figure}
\epsscale{1.0}
\plotone{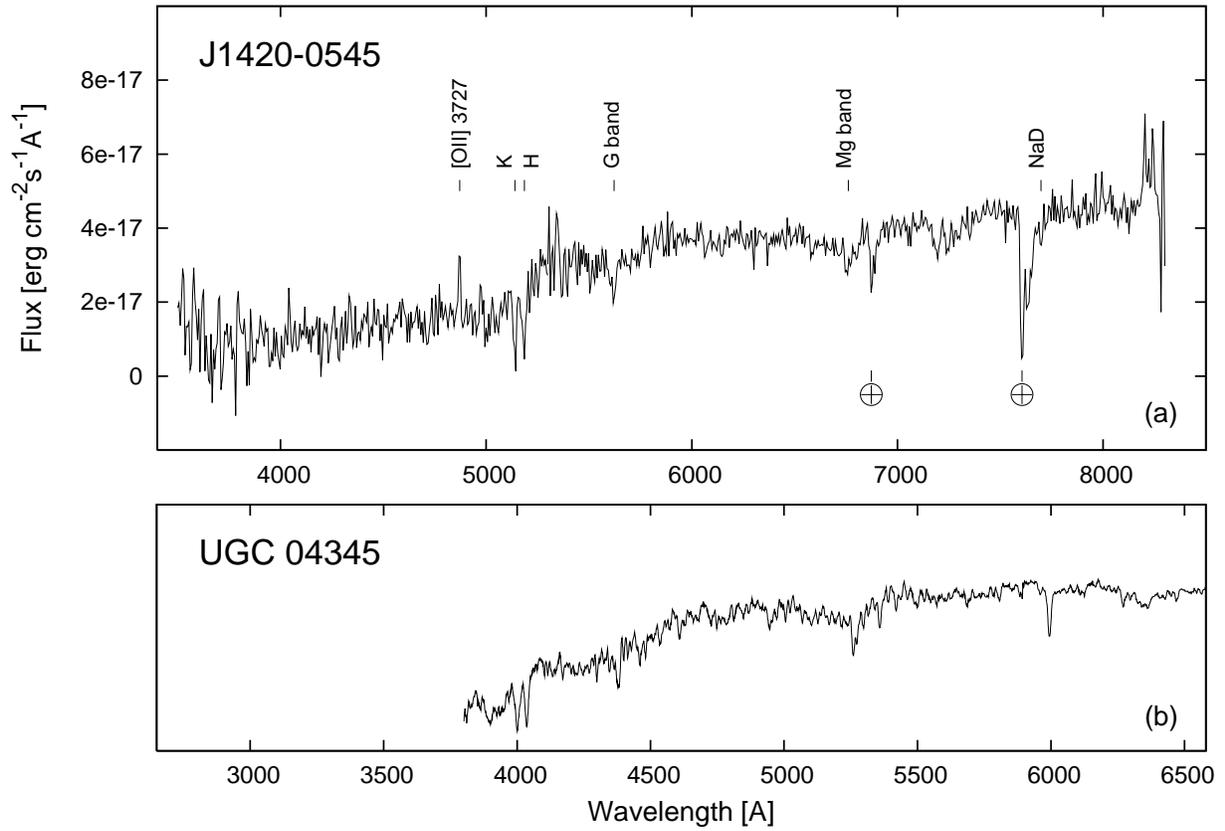}
\caption{{\bf ($a$)} Optical spectrum  of J1420$-$0545.
{\bf ($b$)} Typical spectrum of an early-type radio-quiet galaxy, for comparison}.\label{fig3}
\end{figure}

The obtained spectra were reduced using the standard IRAF {\sc longslit} package. All
the spectra were corrected for bias, flat fields, and cosmic rays, as well as being wavelength- and
flux density-calibrated. The wavelength calibration was performed by using exposures to CuAr/CuNe arc lamps,
wherea the flux calibration was performed by using exposures of the spectrophotometric standard
star SP 1545+035. The standard star was observed directly before and after the target galaxy.
The spectra from each arm were averaged separately and were then combined into a common one--dimensional
spectrum shown in Figure~\ref{fig3}$a$. The spectrum is typical of elliptical galaxies whose
continuum emission with the characteristic 4000\,\AA\hspace{1mm}discontinuity and $H$ and $K$
absorption lines is dominated by evolved stars. The spectrum of the S0-type galaxy
UGC\,04345, as a typical spectrum of an early morphological type galaxy,  is shown in Figure~3$b$
for comparison's sake. A Gaussian profile was fitted to the lines and bands
recognized in the spectrum, and the wavelengths corresponding to the profile centers are used
to calculate the redshift. The emission line and absorption bands detected, as well as the
resulting redshifts of these features, are listed in Table 1.
\noindent
If we average the values in column (3) of Table 1, we find that the redshift of J1420$-$0545 is
$z$=0.3067$\pm$0.0005.

\section{Physical parameters and dynamical age}

Given the spectroscopic redshift, we can determine observational parameters that characterize
the GRG reported; i.e., the linear size, $D$, and the volume of the cocoon, $V_{\rm c}$, via the
axial ratio AR, and radio luminosity, $P_{\nu}$. Other physical parameters are derived by
modeling dynamics of the source with the DYNAGE algorithm \citep{mach}. 
This algorithm is an extension of the analytical model for the evolution of FR type II radio
sources, combining the dynamical model of \citet{ka} with the model for expected radio
emission from a source under the influence of energy loss processes published by \citeauthor{kda} (1997, hereafter KDA). 
The DYNAGE algorithm allows us to determine the values of five of the 
model parameters, i.e., the jet power, $Q_{\rm jet}$, central core density, $\rho_{0}$,
internal pressure in the cocoon, $p_{\rm c}$, dynamical age, $t$, and initial (injected)
spectral index, $\alpha_{\rm inj}$; however, there are only four independent parameters, since
any combination of three parameters from the set $Q_{\rm jet}$, $\rho_{0}$, $p_{\rm c}$,
and $t$ uniquely determines the remaining fourth parameter. The determination of their
values is made possible by the fit to the observational parameters of a source: $D$, AR,
$P_{\nu}$, and the radio spectrum ($\alpha$), which usually provides $P_{\nu}$ at a number of
observing frequencies $\nu=1,2,3$... The values of other free parameters of the model
have to be assumed.

\subsection{Linear size, volume, and luminosity}

The high symmetry of the radio structure and the weak strength of the core suggests that
an inclination angle of the jet axis towards the observer's line of sight is large, possibly
close to 90\degr. Thus, the linear size of J1420$-$0545 is $D$=4690 kpc; i.e. it is larger than
the measurement of 4380 kpc for 3C236, the largest GRG known (with an angular extent of 2430\arcsec\hspace{1mm}and
a redshift of 0.09883). For the dynamical considerations, the volume of the source is needed.
In KDA and DYNAGE a cylindrical geometry of the radio cocoon is assumed; thus,
$V_{\rm c}$=$\pi D^{3}/(4 AR^{2})$. Following \citet{leahy},
we measure the axial ratio AR as the ratio between the angular
size of the total structure and the average of the full width ($w$) of its two lobes. The
latter is determined as the largest deconvolved width of the transversal cross section through
the lobes measured between 3 $\sigma$ total-intensity contours. We find values of $w\approx 390$ kpc and
$AR\approx 12$; however, this latter value is likely an upper limit for the axial ratio, because
the cocoon's width further down towards the core can be larger than the width of the detected
lobes. The dependence of age and other physical parameters of J1420$-$0545 on the value of
AR is considered in the next subsection. 
The 619 MHz and 1400 MHz luminosities are $P_{619}=3.60\times 10^{24}$
W\,Hz$^{-1}$sr$^{-1}$ and $P_{1400}=2.06\times 10^{24}$ W\,Hz$^{-1}$sr$^{-1}$,
respectively. Because of the limited radio data, the model parameters derived
in the next subsection are fitted using the above two luminosity values only.

\subsection{Age and other parameters}

\begin{figure*}
\epsscale{1.0}
\plotone{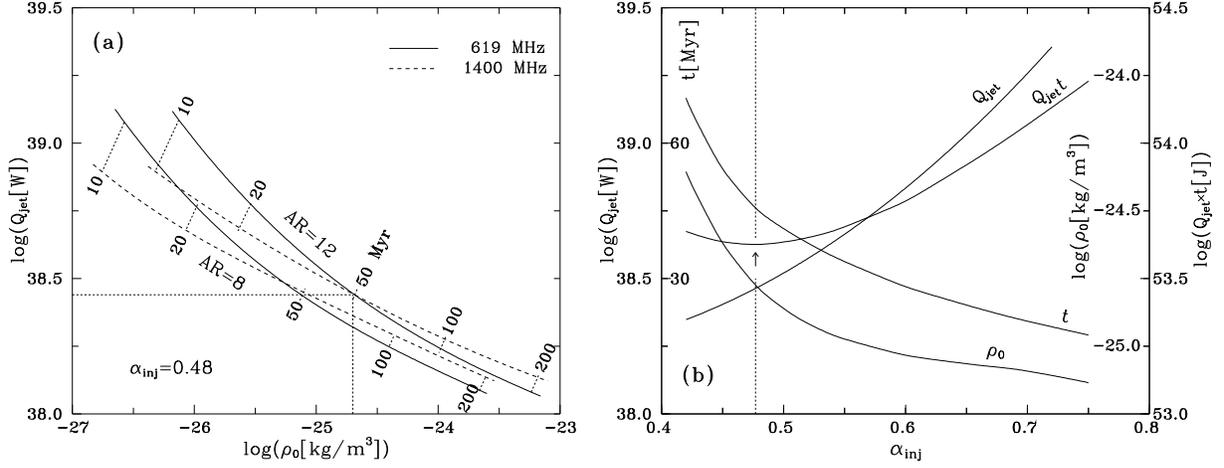}
\caption{{\bf ($a$)} Jet power vs. central core density ($Q_{\rm jet}-\rho_{0}$) diagram at the two observing frequencies,
for $\alpha_{\rm inj}$=0.48 and two different values of AR. An intersection
of the solid and dashed curves indicates an age solution and gives the corresponding values
of $Q_{\rm jet}$ and $\rho_{0}$ for a given value of $\alpha_{\rm inj}$ value. The dotted
lines indicate the age to which the fitted values of $Q_{\rm jet}$ and $\rho_{0}$
are related.{\bf ($b$)} Fitted values of age $t$ as well as $Q_{\rm jet}$,
$\rho_{0}$, and $Q_{\rm jet}\times t$, as a function of $\alpha_{\rm inj}$. The
arrow points to a minimum of the product $Q_{\rm jet} t$ (i.e. a minimum of
the total kinetic energy delivered to the cocoon by the jet), and the vertical dotted
line indicates the corresponding value of $\alpha_{\rm inj}$ that determines the
{\sl best} solution for the age.}.\label{fig4}
\end{figure*}

The model requires the density distribution of the ambient medium (which is invariant with redshift)
to be modeled by a power law, $\rho_{\rm a}(r)$=$\rho_{0}(r/a_{0})^{-\beta}$, where $a_{0}$
is the core radius and the exponent $\beta$ describes the density profile. With this
assumption, the age of a radio structure of linear size $D$ is

\begin{displaymath}
t\propto D^{\frac{5-\beta}{3}}\left( \frac{\rho_{0}a_{0}^{\beta}}{Q_{\rm jet}}\right)^{1/3},
\end{displaymath}

\noindent
the pressure within the cocoon is

\begin{displaymath}
p_{\rm c}\propto \left( \rho_{0}a_{0}^{\beta}\right)^{3/(5-\beta)}
Q_{\rm jet}^{(2-\beta)/(5-\beta)}t^{(-4-\beta)/(5-\beta)},
\end{displaymath}

\noindent 
and the energy density and the corresponding equipartition magnetic field strength are

\begin{displaymath}
u_{\rm c}[J\,m^{-3}]=p_{\rm c}/(\Gamma_{\rm c}-1); \hspace{6mm} B_{\rm c}[nT]=
\left(\frac{24}{7}10^{11}\pi\,u_{\rm c}\right)^{1/2},
\end{displaymath}

\noindent
where $\Gamma_{\rm c}$ is the adiabatic index of the lobes' (cocoon) material. If we take
the same values of the model's free parameters  as in \citet{mach}, in particular
$a_{0}$=\,10 kpc, $\beta$\,=\,1.5, and $\Gamma_{\rm c}$=\,5/3, the values of $Q_{\rm jet}$,
$\rho_{0}$, and $p_{\rm c}$ are determined by the fit to the observed luminosities and volume
of the source (cf. KDA). But that fit depends on $\alpha_{\rm inj}$. In the KDA model this also
has to be assumed; however, the DYNAGE algorithm allows us to find its value through a search
for the minimum energy density ($u_{\rm c}$), which, in turn, is proportional to the
minimum kinetic energy delivered to the source (cocoon) by the jets during its actual
age; i.e., 2$Q_{\rm jet}\,t$. The fitted values of $Q_{\rm jet}$ and $\rho_{0}$ for
$\alpha_{\rm inj}=0.48$ and different hypothetical ages of J1420$-$0545 are shown in the
log\,$Q_{\rm jet}$--log\,$\rho_{0}$ diagram in Figure~\ref{fig4}$a$. The two pairs of
curves (solid and dashed) give the fits for two different values of AR; the
one determined from the width of the observed lobes (cf. \S 4.1), and the
same, but decreased by a factor of 1.5. This figure shows that both pairs of
curves corresponding to two different observing frequencies intersect each other
at a similar age of 45--50 Myr. Thus, by enlarging the cocoon's volume by 
a factor of $(1.5)^{2}$, we find a significant decrease of the core density $\rho_{0}$
compared with that for AR=12, while the values of the age $t$ and
$Q_{\rm jet}$ remain comparable.

The dependence of the resulting model solutions of $t$, $Q_{\rm jet}$,
$\rho_{0}$, and $Q_{\rm jet}t$ on $\alpha_{\rm inj}$ for AR=12 is
shown in Figure~\ref{fig4}$b$. The vertical line and the arrow indicate the value of 
$\alpha_{\rm inj}$ that corresponds to the minimum value of the product
$Q_{\rm jet} t$. The minimum of this product, which gives the kinetic
energy delivered to the cocoon by the jet, corresponds to a minimum of the energy
density in the cocoon, $u_{\rm c}$, as is expected from the KDA model assumption
of energy equipartition between magnetic fields and particles. The parameter values
resulting from the fit, as well as their errors, are listed in Table~2  
and are compared with the corresponding values derived for 3C236 (J. Machalski 2008, in preparation).
The errors are calculated by means of uncertainties of the parameters ($a, b, c$)
in the analytical forms $y=a+bx+cx^{2}$ or $y=a+bx+c\,exp(\pm x)$, which are used to
smooth the data points in the plots of $y$ versus $x$, where $x\equiv \alpha_{\rm inj}$
and $y$ is equivalent to either $t$, log($Q_{\rm jet}$), or log($Q_{\rm jet} t$)
(cf. Figure~\ref{fig4}$b$).
These uncertainties were determined during the fitting procedure of a set of
pairs ($x$, $y$) with the above forms. Therefore, the errors given in Table~2
do not account for uncertainties of the model's free parameters, such as $a_{0}$,
$\beta$, $\Gamma_{\rm c}$, etc., as well as that of the observational parameter AR.
A possible influence of uncertainties of the above parameters on the DYNAGE predictions
has been discussed in \citet{mach}.
Because of limited radio spectral data for J1420$-$0545, we suppose
that the quoted errors for this radio galaxy may be larger by a factor of 2.

\subsection{Discussion of the results}

The DYNAGE model solution for J1420$-$0545 strongly suggests that this radio
galaxy is at least twice as young as other GRGs at redshifts of about 0.1,
and that it evolved in an exceptionally low-density environment. The fitted value
of $\rho_{0}$ is a factor of 100 lower than the central densities of
typical host galaxies themselves, (e.g. \citealt{canizares}; \citealt{mulchaey}; \citealt
{hardcastle}) and lower than the halo densities  in clusters of galaxies,
(e.g. \citealt{jones}; \citealt{schindler}). In particular, the value of $\rho_{0}$
found for J1420$-$0545 is about 20 times lower than that for 3C236
(cf. Table~2). Enlarging the cocoon's width, i.e. assuming that AR=8,
causes a further decrease of the fitted value of $\rho_{0}$ by a factor of
3.4. On the other hand, the above enlargement does not change significantly
the model solutions for the values of $\alpha_{\rm inj}$ and $t$. For either
value of AR, the minimum of the jets' kinetic energy is at an $\alpha_{\rm inj}$
value of around 0.48, and the age estimate is around 45 Myr. Such a relatively
young age seems to be confirmed by evident hot spots that were clearly visible in both of the
lobes of the source when it was mapped using the FIRST survey.

Having modeled the values of age and jet power with DYNAGE, we can estimate
the value of the IGM (ambient) density at the jet heads, $\rho_{\rm a}$. Balancing the
ram pressure of the IGM and the thrust of the jet, \citet{begel} derived the
equation for the age of a FR type II source (their eq. [3]),

\begin{displaymath}
t\approx \left( \frac{\rho_{\rm a}}{Q_{\rm jet}v_{\rm jet}A_{\rm h}}
\right)^{1/4}A_{\rm c},
\end{displaymath}

\noindent
where $A_{\rm h}$ and $A_{\rm c}$ are the cross-sectional areas of the bow shock
at the head of the jet and of the cocoon, respectively. Thus, if we assume that
the jet speed is close to the speed of light and take into account that the areas
$A_{\rm h}$ and $A_{\rm c}$ can be derived from the radio maps, a value of
$\rho_{\rm a}$ can be calculated from the above equation.
Usually $A_{\rm h}$ is identified with the cross-sectional area of hot spots. The
deconvolved lateral angular sizes of the hot spots in the lobes of J1420$-$0545
are 3.8$\arcsec$ and 4.1$\arcsec$. Averaging these values, we find that their mean
diameter is 17.7 kpc; hence, $A_{\rm h}$=$2.34\times 10^{41}$ m$^{2}$, whereas
$A_{\rm c}$=$1.14\times 10^{44}$ m$^{2}$ for AR=12. Substituting these values
of $A_{\rm h}$ and $A_{\rm c}$, as well as the values of $t$ and $Q_{\rm jet}$
from Table~2, into the above equation, we find that $\rho_{\rm a}$=$5.9\times 10^{-28}$
kg\,m$^{-3}$. If the cocoon were (1.5)$^{2}$ times fatter (i.e. if AR=8), the
resulting ambient density would be lower by a factor of about 37 and would
reach an unacceptably low value of about $1.6\times 10^{-29}$ kg\,m$^{-3}$.
Such a low value of AR is also very unlikely because the ambient density of
$1.6\times 10^{-29}$ kg\,m$^{-3}$ would correspond to that calculated from the
density profile $\rho_{\rm a}(r_{\rm h})$=$\rho_{0}(r_{\rm h}/a_{0})^{-\beta}$
at a radius of $r_{\rm h}\approx 2.2$ Mpc! A power-law density distribution at
such a very large distance from the host galaxy is absolutely improbable.
On the other hand, the value of $5.9\times 10^{-28}$ kg\,m$^{-3}$ seems to be
reasonable as it corresponds to a halo radius of $r_{\rm h}\approx
445$ kpc. This radius is well within the range of radii estimated for clusters
of galaxies by \citet{jones}. 

There is also the possibility that the gaseous halo of the host galaxy is much
smaller, and although the halo's density profile in the case of
J1420$-$0545 is not known, its jets may propagate through the (nearly) uniform IGM for a
considerable fraction of the entire source lifetime. Therefore, while assuming
that the source dynamics has been governed by the uniform environment, i.e.
$\beta$=0, leads to an unphysical picture that the density at the
head of jets, 2.2 Mpc away from the host galaxy, would be equal to its central
density, it may be worthwhile to check how the fitted parameters of the source,
i.e., the density in the front of the jets, $\rho_{\rm a}$, the jets' power, and the
source age, would change if $\beta$ was equal to zero (still for $AR$=12).

As a result, we find that $Q_{\rm jet}$ would be almost twice as strong as its
value in Table~2, and the ambient density $\rho_{\rm a}$ would be lower by a
factor of 4 than $5.9\times 10^{-28}$ kg\,m$^{-3}$. Moreover, the source
would be also younger than 47 Myr, which, in turn, would imply some even higher average
expansion velocity of the source along the jet axis than the value provided in Table~2.
The fitted values of the above parameters are given in Table 3. Certainly, an
average expansion velocity of about 0.21\,$c$ (even 0.163\,$c$) is terribly
higher than its values inferred from the spectral-ageing analyses for extended
FR type II radio sources (e.g. \citealt{alex}; \citealt{liu})
and for FR type II GRGs \citep{jamrozy}, but it is rather similar to the advance
speeds of sources not larger than 10--20 kpc (cf. \citealt{murgia}). Although
the dynamical age of lobes can be significantly higher than the
synchrotron age of the oldest radiating particles (cf. \citealt{mach}), such a
high advance speed of J1420$-$0545 is quite an unexpected and puzzling result, because if
the external density decreases more slowly than $r^{-2}$  (where $r$ is the
distance from the core; cf. \S~4.2), which is the necessary condition for forming the head
of the jet and to observe a source of FR type II, the advance speed does decrease with
the source's age (e.g. \citealt{falle}). Thus the above result may imply that
the head of the jets in our galaxy must have propagated with a speed characteristic
for very young compact double sources, or even comparable to the speed of light,
not too long ago. This is evidently related to the very low density of the IGM
found in the modelling.

There are two possible explanations that could account for lowering the true age of the
source and enhancing its expansion speed:

1. Large errors in the flux densities used to calculate the radio
luminosity of the lobes (cocoon) at the two observing frequencies. If the radio
spectrum determined from these flux densities is flatter than the actual one, the
source will appear younger in the modelling procedure; 

2. Not all the kinetic energy in the jet flow had been transferred to the magnetic
field and the relativistic electrons in the lobes. Such a situation is actually
foreseen in the KDA model, in which the ratio ($k'$) of the energy densities of
thermal particles to that of electrons when they are spread into the cocoon
during the injection is defined. 

\noindent
In order to provide a quantitative check of the possible influence of the uncertain radio spectrum, we
generated a fiducial spectrum steeper than that derived from the observations available to date
and ran the fitting procedure for the frequency range 150--4800
MHz. We found that the age of J1420$-$0545 could possibly rise
to no more than 60$\pm$7 Myr.

Following \citet{kda}, up to this point we have neglected any influence of the thermal particles in the
lobes of J1420$-$0545; i.e. $k'$=0 was assumed. However, in order
to check how the thermal plasma can modify the model parameters derived from the
fit, we performed comparison calculations in which we set $k'$=1 and $k'$=10. The
results are given in the last two rows of Table~3. It appears that all the
parameters $Q_{\rm jet}$, $\rho_{0}$, $\rho_{\rm a}$, $u_{\rm c}$, $p_{\rm c}$, and $t$ do increase with
$k'$, while, on the other hand, an increase of $k'$ causes a decrease of the advance speed of the lobes.
We note that the age $t$ increases in a much slower manner than do the other parameters.
Thus, a very high value of $k'$ (k=$k'$$\gg$10) would be necessary to reduce the
advance speed of the galaxy to 0.01$c$--0.04$c$ (i.e. the values typical for
the extended double radio sources; cf. \citealt{brw}), but speeds as high as 0.10\,$c$--0.15\,$c$
are not ruled out on the basis of the observed asymmetries of sources \citep{scheuer}.
Also, the jets may not keep expanding with a constant opening angle and it is not at all
clear that the velocities will keep systematically decreasing. The hot spot size seems
to increase with the overall source size untill about 10--20 kpc, after which it appears
to be more constant (cf. \citealt{js}). In that paper the authors listed sources that had 
observational evidence of recolimation of the jets. In such instances, the hot spot
size would be smaller, which would increase the pressure due to the beam in the hot spots.
It may also be important perhaps that J1420$-$0545 has significant hot spots, unlike most of
the known giants, whose lobe structures are relatively diffused.

To summarize, we argue that the fitted parameters for J1420$-$0545 as given in
Table~2 seem to be quite realistic ones, and the current uncertainty of its radio
spectrum and the unknown influence of thermal plasma are not likely to significantly
change the values derived in this paper. The very low ambient density around the
lobes of our GRG strongly supports a hypothesis of large heterogeneity of the
IGM.

\subsection{Comparison to other radio galaxies of different linear size}

\begin{figure*}
\epsscale{1.0}
\plotone{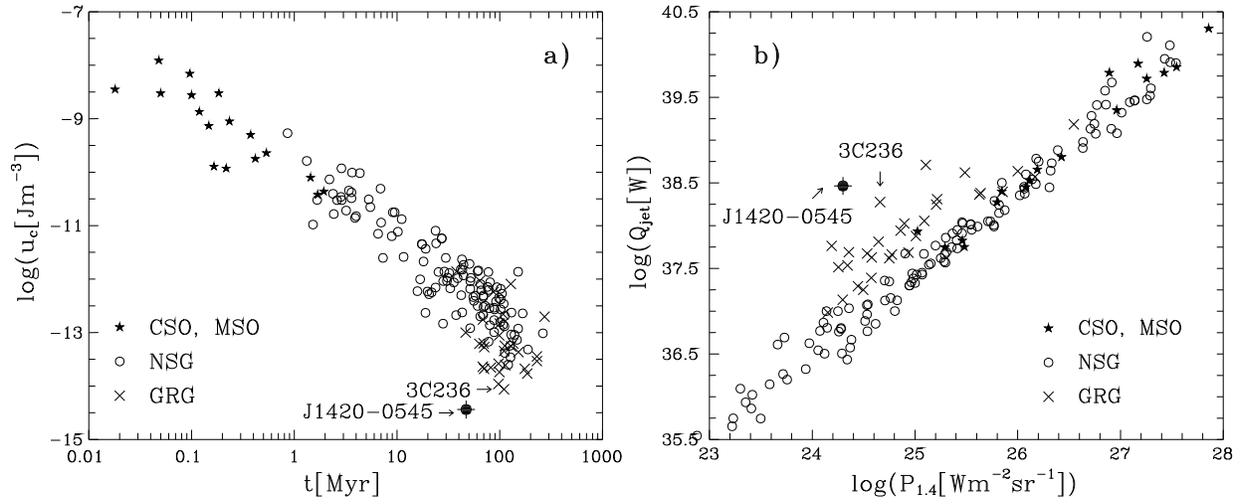}
\caption{Loci of J1420$-$0545 on the diagrams of {\bf $a$)} energy density of the source vs.
its age and {\bf $b$)} the jet power vs. the 1.4 GHz luminosity, comprising FR type II radio galaxies
of different linear sizes}.\label{fig5}
\end{figure*}

In order to compare the derived parameters of J1420$-$0545 with the corresponding
parameters of other FR type II radio galaxies we use the sample of 30 GRGs and 120
normal-sized galaxies (NSGs) analyzed by \citet{mj}, 
for which dynamical
age is determined with the DYNAGE. These data are supplemented with 17 small-sized double
radio galaxies of the CSO and MSO types (compact and medium symmetric objects) for which
the DYNAGE algorithm has given satisfactory results (J. Machalski 2008, in preparation).
Figure~\ref{fig5} shows
how the estimated parameters of J1420$-$0545 differ from the relevant parameters of other
sources used for comparison. For example, the energy density of its lobes/cocoon,
$u_{\rm c}$, as well as the corresponding pressure, $p_{\rm c}$, are extremely low
(Figure~\ref{fig5}$a$). The value of $p_{\rm c}$ (in Table 2) is a factor of 2.5 lower
than that for 3C236, although the redshifts of J1420$-$0545 and 3C236
differ by a factor of 3.  This confirms that J1420$-$0545 actually traces very low
density environments. The ambient density at the head of the lobes, calculated from
the dynamical arguments in \S4.3 corresponds to a particle (proton) density of
$\sim 10^{-7}$ m$_{\rm p}$\,cm$^{-3}$.

Furthermore, Figure~\ref{fig5}$b$ shows that GRGs are characterized by much lower
radio luminosities than are NSGs of the same jet power.
This is what one expects using the KDA model. The model predicts that the
luminosity in the inverse-Compton--dominated regime decreases as
$D^{-(4+\beta)(\frac{1}{3}+\Gamma_{\rm B})/3\Gamma_{\rm c}}$ or
$t^{-(4+\beta)(\frac{1}{3}+\Gamma_{\rm B})/\Gamma_{\rm c}(5-\beta)}$, where
$\Gamma_{\rm B}$ is the adiabatic index of a magnetic "fluid" and is equal to 4/3.
Thus, for the assumed value of $\beta$=1.5 and $\Gamma_{\rm c}$=5/3, one finds that
$P_{\nu}\propto D^{-11/6}$ or $P_{\nu}\propto t^{-11/7}$. Figure~\ref{fig5}$b$
shows that J1420$-$0545 is less luminous than NSGs with a comparable jet power by a factor of about 100. 
Applying the predicted luminosity
evolution with size $D$, one would expect J1420$-$0545 to have had the same
luminosity as those NSGs when its linear size was about 380 kpc. On the other
hand, a similar approach to the luminosity evolution with age $t$ suggests that
the investigated radio galaxy would have had a luminosity that was higher by a factor of 100 when
it was 2.5 Myr old. However, among 167 sources in the sample included in
Figure~\ref{fig5}, there are none that are both younger than 10 Myr and have a size larger
than 200 kpc. It can be calculated that in this sample there is no source with
$\rho_{0}<6\times 10^{-25}$ kg\,m$^{-3}$ and only three with
$\rho_{0}\leq 10^{-24}$ kg\,m$^{-3}$. Therefore, although the largest radio
galaxies can likely have evolved from younger and smaller sources, J1420$-$0545 is
evidently a nontypical GRG due to its exceptionally low-density (a void?)
environment as found by the fit.

\acknowledgements

The authors thank the Service Programme staff for the WHT observations, Alex Kraus 
for his help with the Effelsberg observations, and Michal Siwak 
for the Mount Suhora observations, as well as the Centre Director of the NCRA, TIFR, for
the allocation of discretionary observing time and Chiranjib Konar for his assistance.
The authors appreciate the comments and helpful 
suggestions of the anonymous referee. The Digitized Sky Surveys were produced at
the Space Telescope Science Institute under US government grant NAG W-2166.
The images of these surveys are based on photographic data obtained using the Oschin
Schmidt Telescope on Palomar Mountain and the UK Schmidt Telescope. This work was
supported in part by the Ministry of Science and Higher Education of Poland (MNiSW) with funds for 
scientific research
in 2005-2007 under contract 5425/PO3/2005/29. D.K.-W. was partly 
supported by MNiSW funds for scientific research in 2007-2008 
under contract 4272/H03/2007/32.

\clearpage
\begin{table}[h]
\begin{center}
\caption{Spectral Features Detected and Their Redshifts}
\begin{tabular}{lcc}
\hline
Line/Absorptione Band Detected  & $\lambda_{\rm obs}$  & $z_{\lambda}$ \\
          & (\AA) & \\
\hline
[$O$ II] $\lambda$3727         & 4870.9  & 0.3069 \\
$Ca$ II $\lambda$3934         & 5142.1  & 0.3071 \\
$Ca$ II $\lambda$3968         & 5185.6  & 0.3069 \\
$G$ band $\lambda$4305       & 5622.0   & 0.3059 \\
Mg band $\lambda$5175      & 6762.0   & 0.3067 \\
\hline
\end{tabular}
\end{center}
\end{table}

\begin{table}[h]  
\begin{center}
\caption{Physical Parameters of J1420$-$0545 (for AR=12) and 3C236 (AR=16)}
\begin{tabular}{llll}
\hline
Parameter   & Symbol/Relation         &  Value for J1420$-$0545 & Value for 3C236\\
&&& \\
%(1)            & (2)                        & (3) & (4)\\
\hline
Initial effective\\
spectral index   & $\alpha_{\rm inj}$    & 0.477$\pm$0.007  & 0.486$\pm$0.009\\
Age              & $t$\,[Myr]            & 47$\pm$6         & 110$\pm$18\\
Jet power        & $Q_{\rm jet}$\,[W]  &(2.91$\pm$0.30)$\times$10$^{38}$ & (1.88$\pm$0.20)$\times$10$^{38}$\\
Core density     & $\rho_{0}$\,[kg\,m$^{-3}$] &(1.78$\pm$0.35)$\times$10$^{-25}$ & (3.73$\pm$0.45)$\times$10$^{-24}$\\
Kinetic energy   &2$Q_{\rm jet}\,t$\,[J]&(8.50$\pm$0.90)$\times$10$^{53}$ & (6.24$\pm$0.64)$\times$10$^{53}$\\
Energy density& $u_{\rm c}$\,[J\,m$^{-3}$]  & (0.36$\pm$0.02)$\times$10$^{-14}$ & (0.89$\pm$0.05)$\times$10$^{-14}$\\
Internal pressure& $p_{c}$\,[N\,m$^{-2}$] & (2.40$\pm$0.13)$\times$10$^{-15}$ & (5.94$\pm$0.33)$\times$10$^{-15}$\\
Magnetic field   & $B_{c}\,[nT]$    & 0.062$\pm$0.010   & 0.098$\pm$0.014\\
Ratio of energy\\
delivered/radiated &2$Q_{\rm jet}\,t/u_{c}V_{c}$& 14.2$\pm$4.8 & 21.2$\pm$5.1\\
Expansion speed  & $D/(2\,c\,t)$  & 0.163$\pm$0.021         & 0.065$\pm$0.010\\
\hline   
\end{tabular}
\end{center}
\end{table}

\clearpage

\begin{table}
\begin{center}
\caption{The Influence of the Values of $\beta$ and $k'$ on the Basic Physical
Parameters.}
\begin{tabular}{lrcccccc}
\hline
&& $Q_{\rm jet}$ & $\rho_{0}$ & $\rho_{a}$ & $u_{c}$ & $t$ & \\
$\beta$ & $k'$ & (W) & (kg\,m$^{-3})$ & (kg\,m$^{-3}$) & (J\,m$^{-3}$) & (Myr) & $D/(2ct)$\\
\hline
0       &  0   & $5.37\times 10^{38}$ & $1.55\times 10^{-28}$ & $4.20\times 10^{-28}$ & $0.20\times 10^{-14}$ & 37$\pm$3 & 0.207\\
1.5     &  0   & $2.91\times 10^{38}$ & $1.78\times 10^{-25}$ & $5.90\times 10^{-28}$ & $0.36\times 10^{-14}$ & 47$\pm$6 & 0.163\\
     &  1   & $4.04\times 10^{38}$ & $3.00\times 10^{-25}$ & $1.05\times 10^{-27}$ & $0.54\times 10^{-14}$ & 50$\pm$5 & 0.153\\
     &  10  & $8.63\times 10^{38}$ & $1.28\times 10^{-24}$ & $5.66\times 10^{-27}$ & $1.46\times 10^{-14}$ & 63$\pm$8 & 0.121\\
\hline
\end{tabular}
\end{center}
\tablecomments{In the solution for $\beta=k'=0$,  we have $\rho_{0}<\rho_{\rm a}$, 
where all the values of $\rho_{\rm a}$ are calculated from the eq. (3) of \citet{begel}.}
\end{table}
\clearpage

\end{document}